\magnification=\magstep1 
\font\bigbfont=cmbx10 scaled\magstep1
\font\bigifont=cmti10 scaled\magstep1
\font\bigrfont=cmr10 scaled\magstep1
\vsize = 23.5 truecm
\hsize = 15.5 truecm
\hoffset = .2truein
\baselineskip = 14 truept
\overfullrule = 0pt
\parskip = 3 truept
\def\frac#1#2{{#1\over#2}}

\nopagenumbers
\topinsert
\vskip 3.2 truecm
\endinsert
\centerline{\bigbfont ENTANGLEMENT PROPERTIES OF} 
\vskip 8 truept
\centerline{\bigbfont QUANTUM MANY-BODY WAVE FUNCTIONS}
\vskip 20 truept
\centerline{\bigifont J. W. Clark,$^1$ A. Mandilara,$^1$ and 
M. L. Ristig$^2$}  
\vskip 8 truept
\centerline{\bigrfont Department of Physics, Washington University}
\vskip 2 truept
\centerline{\bigrfont St. Louis, Missouri 63130, USA} 
\vskip 8 truept
\centerline{\bigrfont $^2$Institut f\"ur Theoretische Physik, Universit\"at 
zu K\"oln}
\vskip 2 truept
\centerline{\bigrfont D-50937 K\"oln, Germany} 
\vskip 1.8 truecm

\centerline{\bf 1.  INTRODUCTION}
\vskip 12 truept

The quantum information community is currently engaged in a major
effort to quantify the entanglement content of states of multipartite
quantum systems [1-9].  A multipartite system is made up of a number
of parts, which may be identified with individual particles or with 
groups of particles.  Entanglement (actually, {\it Verschr\"ankung} 
is the name given by Schr\"odinger [10] to the nonlocal correlations 
responsible for violations of the Bell inequalities [11].  This 
property has emerged as a physical resource [12,13], analogous 
to energy as a resource for useful work, which is to be drawn upon 
in schemes for quantum communication and quantum computation.  

Since entanglement represents uniquely quantal correlations, it 
becomes of great interest to elucidate the entanglement properties 
of the wave functions commonly used to describe strongly correlated
quantum many-body systems in condensed-matter physics, hadronic
physics, and quantum chemistry.  The information gained in such 
a program should improve our understanding of quantum phase transitions 
occurring in these systems as well as their behavior in regions 
away from critical points.  Here we shall take a first step 
in this direction by quantifying the entanglement of correlated
variational wave functions that have been developed to treat model
systems of interacting Pauli spins localized on the sites
of a regular lattice, i.e., spin-lattice models [14,15].

A pure state of a multipartite quantum system is entangled if and 
only if its state vector is non-separable, meaning that it is not
the direct product of state vectors of the parts.  In many-body language, 
reading ``parts'' as ``particles,'' the wave function cannot be written 
as a product of single-particle wave functions of some basis.  A mixed 
state, which may generally be represented by a density operator, 
is nonseparable and therefore entangled if and only if it is not 
decomposable as a mixture of product states.  

Exchange correlations in Fermi and Bose ground states do not 
contribute to entanglement as a useful resource [16-19]; 
accordingly, a state described by a single Slater determinant or 
permanent is to be considered separable.  Thus, in examining
the entanglement of a many-body wave function, one is in essence
addressing its non-mean-field properties, which reflect fluctuation 
effects due to the presence of strong interactions.  Any subset
of the particles in a system of interacting particles in a
pure state is necessarily in a mixed state.

Bipartite (two-party) entanglement of pure and mixed states 
has received thorough study, especially for the case that 
the two subsystems are two-level systems or Pauli spins.
(This is of course the case of most immediate concern for
quantum computation, where the two-level computing elements
are called qubits.)  While the quantification of bipartite
entanglement is well under control, analysis of multipartite
entanglement quickly becomes a formidable problem as the number 
of parties increases beyond three.  For an $N$-partite
quantum system, $N>2$, entanglement is not characterized by
a single quantity, but rather by a non-unique set of quantities
that grows polynomially with increasing $N$.  Understandably,
there is as yet no consensus on the best choice of such quantities.  

In the present work we will consider only (i) bipartite entanglement 
of a single spin with the rest of the spins in the lattice and
(ii) bipartite entanglement of two spins in the lattice environment.  
Accordingly, our treatment of spin-lattice models will involve 
the following standard measures of bipartite entanglement [20-22]: 
von Neumann entropy, entanglement of formation, concurrence, and 
localizable entanglement.  Considering the transverse Ising model 
laid out on regular lattices (square, cube, tessaract) in two, three, 
and four dimensions, information on these quantities will be gathered 
from available results on the one- and two-site (or one- and two-spin) 
density matrices corresponding to Hartree-Jastrow ground-state wave 
functions [23-25,15].  Where possible, comparison will be made 
with results of earlier work on exactly soluble models [26,27,22] 
or stochastic simulation methods [28]. We also make an interesting
simple connection of the Hartree-Jastrow functions with the
nilpotent polynomial representation of entanglement [9], which 
permits us to expose important qualitative features of these 
trial ground states.

Section II provides the necessary formal and conceptual background
on entanglement measures and their possible role in identifying
and characterizing quantum phase transitions.  In Section III 
we introduce the transverse Ising model and sketch its analysis 
and treatment within the framework of correlated-basis theory and 
hypernetted-chain techniques.  The numerical results and attendant 
discussion are presented in Section IV.  In Section V we look ahead 
to more ambitious explorations of the entanglement properties 
of the many-body wave functions employed in correlated-basis 
and coupled-cluster approaches to strongly correlated quantum 
systems.
\vfill\eject

\centerline{\bf 2.  ENTANGLEMENT AND QUANTUM PHASE TRANSITIONS}
\vskip 12 truept

The entanglement of the two parts (1,2) of a bipartite system in a
pure state $\rho = |\psi \rangle \langle \psi | $ may be defined as the
{\sl von Neumann entropy}
$$
S = - {\rm tr}(\rho_i \log_2 \rho_i)
 = - {\rm tr}(\rho_j \log_2 \rho_j) \eqno(1)
$$
of either subsystem, where $\rho_i = {\rm tr}_j(\rho)$ 
and $\rho_j = {\rm tr}_i(\rho)$.  When either subsystem is a spin-1/2 system, 
$S$ ranges from 0 (not entangled) to 1 (maximally entangled).

The entanglement between two parts of a system in a mixed state $\rho$ 
(e.g., two spins within a multispin system) is not uniquely defined.  
One natural definition is the {\sl entanglement of formation}, which is 
the minimum, over all pure-state decompositions of $\rho$, of the 
expected entanglement required to construct $\rho$ from such a 
decomposition, using $S$ as a measure of the pure-state entanglement.  
In general, this quantity is awkward to calculate.  However, for 
the case of two spins 1/2 (2 qubits), it can easily be found [21] 
from their density matrix as a simple monotonic function of the 
concurrence $C$,
$$
E_F(\rho) = h \left( {{1 + \sqrt{1-C^2} } \over 2}\right) \eqno(2)
$$ 
with
$$
h(x) = -x \log_2 x - (1-x) \log_2 (1-x)\,. \eqno(3)
$$
 
Given the two-spin density matrix $\rho_{ij}$ obtained by
tracing out all spins other than $i$ and $j$, the
concurrence is calculated as
$$
C(\rho_{ij}) = {\rm max}[0, \mu_1 - \mu_2 - \mu_3
- \mu_4 ] \,, \eqno(4)
$$
where the $\mu_i$ are the eigenvalues (in decreasing order, each real and 
nonnegative) of the Hermitian matrix
$$
R = \left[ \rho^{1/2} {\tilde \rho} \rho^{1/2} \right]^{1/2} \eqno(5)
$$
and
$$
{\tilde \rho} = ( \sigma^y \otimes \sigma^y ) \rho^*
(\sigma^y \otimes \sigma^y ) \eqno(6)
$$
is the spin-flipped density matrix ($\rho_{ij}$ being abbreviated as 
$\rho$).  The concurrence $C$ ranges from zero for a separable state 
to unity for a maximally entangled state.  For a pure state of 
qubits, $ | \psi \rangle = a |00 \rangle + b | 01 \rangle + 
c | 10 \rangle + d | 11 \rangle $, one obtains $C = |ad-bc|$, which 
is clearly a measure of the departure from a product state.  

Other measures of bipartite entanglement of mixed states have
also been proposed.  The {\sl entanglement of assistance} is the 
maximal two-party (e.g., two-spin) entanglement that can be achieved 
by performing any kind of measurement on the other parts of a 
multipartite (e.g., multispin) system.  In a sense, this measure lies 
at the opposite extreme from the entanglement of formation, and 
again it is hard to calculate.  

Verstraete et al.\ and Popp et al.\ [22] have proposed a similar 
measure that is more accessible.  The {\sl localizable entanglement} 
is the maximal amount of entanglement between two parties 
that can, on average, be created -- localized -- by performing 
only {\it local} measurements on the other parts of the system.
Unlike the other bipartite entanglement measures mentioned, it is not
(in general) determined from a knowledge of two-particle correlation 
functions alone.  On the other hand, it captures more complicated features 
of the state of a multipartite system and leads to a natural definition of
entanglement length.  Moreover, although difficult to calculate in
general, one can obtain {\it bounds} (usually tight ones) on its 
magnitude from the connected two-party correlation function.
An upper bound is given by the entanglement of assistance as
measured by its concurrence $C_A$, while a lower bound is provided 
by the maximal connected (or ``classical'') correlation function 
(see Ref.~[22] for details).  

Our study of the entanglement properties of correlated wave functions
for spin lattices was motivated by the work of Osterloh et al.\ [26] and
more especially that of Osborne and Nielsen (ON) [27].  These authors were
the first to explore possible connections between quantum phase
transitions and entanglement.  Both investigations focused on
the anisotropic XY model on a one-dimensional (1D) lattice with
$N$ sites occupied by Pauli spins with nearest-neighbor ferromagnetic
couplings, subject to a transverse magnetic field.  
Since this model is exactly soluble using the Jordan-Wigner transform, 
it admits an incisive analysis of the behavior of entanglement in
the vicinity of a simple quantum phase transition from paramagnetic
disorder to ferromagnetic order.   A special case, the transverse 
Ising model, received the most attention.  There was much subsequent
work on quantum spin chains along similar lines, driven by their
tractability and by the equivalence of spin-1/2 with the qubit
of quantum information theory.  

A {\sl quantum phase transition} is associated with a qualitative change
of the ground state of a quantum many-body system as some parameter
(e.g, density, pressure, doping, coupling constant) is varied.
In contrast to ordinary phase transitions driven by thermal fluctuations
at finite temperature, quantum phase transitions are driven by 
purely quantal fluctuations and can occur at zero temperature.
At the critical point in parameter space where the transition
takes place, long-range correlations develop in the ground state.
Osterloh et al.\ and ON proposed that there must exist an intimate 
relation between quantum phase transitions and entanglement, 
and that the behavior of a suitable entanglement measure should 
bear a signature of the singular behavior of the system near 
the critical point.  Their results generally support this view, 
although some unexpected features were encountered.  For example, 
the maximum of the nearest-neighbor concurrence does not occur 
exactly at the critical point, but at a slightly lower value 
of ${\bar \lambda}$.  Importantly, these studies indicate that
one cannot establish a universal connection between bipartite
entanglement and quantum critical points, but rather that
multipartite measures are necessarily involved in a rigorous
analysis.

Let us consider the transverse Ising model in the form studied by 
ON:
$$
H = - \sum_{j=0}^{N-1} \left({\bar \lambda} \sigma_j^x \sigma_{j+1}^x
+ \sigma_j^z  \right) \,.\eqno(7)
$$
In this form, the constants defining the model are lumped into
a single coupling parameter ${\bar \lambda}$.  ON examined the
entanglement properties of both the ground state at zero temperature
and the thermal mixed state at finite temperature $T$, observing
that the ground state has a two-fold degeneracy which is generally
broken.  We shall be concerned only with their results for the
ground state, whose bipartite entanglement content they measured
in terms of (i) the von Neumann entropy $S$ of the one-site
reduced density matrix and (ii) the concurrence between two spins,
calculated from the two-site reduced density matrix.  In the
first case the two parties are a single spin $i$j and the $N-1$ spins making
up the rest of the lattice system; in the second they are 
two spins $ij$, residing in a mixed state within the lattice system
of the remaining $N-2$ spins.

\noindent
{\sl Single-Site Entanglement}.  The one-site reduced density matrix
for a spin $i$ at an arbitrary site (all being equivalent by translational
invariance) is
$$
\rho_i = {\rm tr}_i(\rho) = {1\over 2}
\sum_{\alpha = 0}^3 q_\alpha \sigma_i^\alpha \,, \eqno(8)
$$
where $\sigma^0 = 1$ and $\alpha=1$, 2, 3 are $x,y,z$, and
$$
q_\alpha = {\rm tr}(\sigma_i^\alpha \rho) =
\langle \sigma_i^\alpha \rangle \,. \eqno(9)
$$
If the full symmetries of the Hamiltonian $H$ are enforced, the number of
terms reduces to just one ($\alpha = 1$).  However, the degeneracy of
the ground state leads to broken symmetry with $q_3 \neq 0$. The two 
parameters required to specify the single-site density matrix in 
the ground state are the longitudinal ($x$ component) and transverse 
($z$ component) magnetizations in either of the two degenerate states, say
$$
M_x = \langle 0^+ | \sigma_x | 0^+ \rangle \,, \qquad
M_z = \langle 0^+ | \sigma_z | 0^+ \rangle \,. \eqno(10)
$$
Thus we have
$$
\rho_i = {1 \over 2} \left( I + M_x \sigma_x + M_z \sigma_z \right)\,, \eqno(11)
$$
and the entanglement as given by von Neumann entropy is
$$
S \equiv - {\rm tr}(\rho_i \log_2 \rho_i )\,, \eqno(12)
$$
with $0 \leq S \leq 1 $.  The two eigenvalues of $\rho_i$ 
are easily found to be 
$$
\kappa_{1,2} = {1 \over 2 } \left(1 \pm \sqrt{M_x^2 + M_z^2} \right) \,,   \eqno(13)
$$
leading to the result
$$ 
\eqalignno{
S &= - \kappa_1 \log_2 \kappa_1  - \kappa_2 \log_2 \kappa_2   \cr
   &=  - {1 \over 2 } \left[ (1+x)\log_2 (1+x) + (1-x) \log_2 (1-x)  \right] 
 & (14) \cr }
$$
for the von Neumann entropy of a single spin with respect to the
rest of the lattice, where
$$
x^2 = M_x^2 + M_z^2 \,. \eqno(15)
$$

\noindent
{\sl Two-Site Entanglement}.  For the two-site reduced density matrix,
similar arguments lead to 
$$
\rho_{ij} = {\rm tr}_{ij}(\rho) = {1 \over 4}
\sum_{\alpha, \beta = 0 }^3 p_{\alpha \beta } \sigma_i^\alpha
\otimes \sigma_j^\beta \,, \eqno(16)
$$
with coefficients
$$
p_{\alpha \beta } = {\rm tr} ( \sigma_i^\alpha \sigma_j^\beta
                     \rho)= \langle \sigma_i^\alpha
                     \sigma_j^\beta \rangle \,, \eqno(17)
$$
and to the expression
$$
\rho_{ij} = {1 \over 4} \left[ I + M_z (\sigma_i^z + \sigma_j^z )
+ \sum_{\alpha = 1}^3 \langle \sigma_i^\alpha \sigma_j^\alpha \rangle
\sigma_i^\alpha  \sigma_j^\alpha \right]  \eqno(18)
$$
in terms of the transverse magnetization $M_z$ and the two-spin 
correlation functions $\langle \sigma_i^x \sigma_j^x \rangle$,
$\langle \sigma_i^y \sigma_j^y \rangle$, and  
$ \langle \sigma_i^z \sigma_j^z \rangle$.  Knowing these ingredients
from an exact solution or approximate many-body treatment, the 
concurrence $C$ may then be determined from Eq.~(4).
\vskip 28truept

\centerline{\bf 3.  VARIATIONAL THEORY OF TRANSVERSE ISING MODEL} 
\vskip 12 truept

We now review the variational-CBF approach [23-25] to the ground state and
elementary excitations of the transverse Ising model in $D$ dimensions.
Here, ``variational'' means that a variational {\it Ansatz} is made for
the ground-state wave function; ``CBF'' means ``correlated basis
functions,'' implying that both the ground-state and excited-state
descriptors will contain nontrivial correlations beyond mean-field
theory.  Here we are only interested in the results for the 
magnetizations and correlation functions in the ground state required
for the evaluation of the relevant bipartite entanglement measures.

Written for arbitrary dimension $D$, the Hamiltonian is
written with a more general parametrization than that employed in
Refs.\ [26,27,22], namely
$$
{\cal H} = {1\over 2} \sum_{i,j}^{N}\Delta_{ij}\sigma_i^x \sigma_j^x 
+ \lambda \sum_i^N (1 -\sigma_i^z)\,. \eqno(19)
$$
The $N$ spins are situated on the lattice sites of a $D$-dimensional 
hypercube.  A generic vector from one site to another will be denoted by
${\bf n}$.  The spin-spin interaction is of the Ising type: 
$$
\Delta ( {\bf n} ) = \cases{ 2D & $ {\bf n} = 0 $ \cr
-1   & for nearest neighbors  \cr
\phantom{-} 0  & otherwise  \cr}   \eqno(20)
$$
with $\Delta_{ij}\equiv \Delta({\bf r}_i - {\bf r}_j)=\Delta({\bf n})$.
The strength of the external field is measured by the coupling parameter 
$\lambda$ ($0 \le \lambda \le \infty$).  For 
the special case $D=1$ (which we do not treat numerically), 
identification of $\lambda$ with ${\bar \lambda}$ brings the
Hamiltonian (19)-(20) into coincidence with the form (7) 
used by Osborne and Nielsen, apart from an overall constant
factor and a constant shift of energy.

CBF theory provides a comprehensive framework for {\it ab initio} microscopic 
description of strongly interacting many-body systems [29].  In application to 
the transverse Ising model, one would like to achieve such a description
for values of the coupling parameter $\lambda$ over its full range
from 0 (corresponding to the strong-coupling limit) to $\infty$
(weak-coupling limit).  Gross properties to be determined include
the longitudinal magnetization $M_x = \langle \sigma_i^x \rangle $
in the normalized ground state, the transverse magnetization
$M_z = \langle \sigma_i^z \rangle$, the spin-spin spatial distribution
function $g({\bf n}) = \langle \sigma_i^x \sigma_j^x \rangle$, and the
corresponding structure function $S({\bf k})$, all in the
ground state.  Further, one would like to determine the ground-state 
energy $E_0$ and the coupling parameter $\lambda_c$ at the quantum 
critical point, where the system changes phase from paramagnetic 
to ferromagnetic (or vice versa).  (In general one would also 
like to find the properties of the elementary excitations,
including the dispersion law and magnon energies.)

To separate the mean-field effects from the effects of dynamical
correlations, it is convenient to introduce a modified (``connected'') 
distribution function
$$
G({\bf n})= (1-M_x^{2})^{-1}\Big[ g({\bf n})-{\delta}_{{\bf n 0}}-
(1-{\delta}_{{\bf n 0}})M_x^{2}\Big] \eqno(21)
$$
and extract the so-called {\sl spin-exchange strength} from $M_z$:
$$
n_{12}={(1-M_x^{2})}^{-{1\over 2}}M_z \> \eqno(22)
$$
In mean-field approximation, $G({\bf n}) \equiv 0$ and $n_{12} \equiv 1$.

The following steps have been taken in the CBF analysis of the
transverse Ising model, and corresponding numerical results are
available [23-25]:
\item{(i)}
Expression of the ground-state energy as a closed functional 
of the longitudinal magnetization $M_x$, the modified
distribution function $G({\bf n})$, and the spin-exchange strength 
$n_{12}$.
\item{(ii)}
Construction of a variational ground state having the
essential correlation structure.
\item{(iii)}
Evaluation of the spatial distribution function and spin-exchange
strength, and hence the energy functional, for a generic trial
ground state.
\item{(iv)}
Optimization of the trial ground state -- derivation and
solution of Euler-Lagrange equations.
\item{(v)}
Evaluation of the desired gross properties and correlation measures,
for the optimal ground state.
\item{(vi)}
Construction of the excited states and associated energies, in
Feynman approximation.

\noindent
Further steps have been envisioned but not carried out:
\item{(vii)}
Systematic improvement of the zero-temperature description, by
inclusion of higher-spin correlations and backflow in
ground-state trial function and excitation {\it Ansatz},
and/or perturbation theory in a basis of correlated
states.
\item{(viii)}
Extension to finite $T$ via correlated density matrix theory [30].

For the purpose of the present work, only the first two steps, (i)
and (ii), require more explicit presentation.  The expression for
the energy functional, applicable to a generic proposal for the
ground state, is given by
$$
{E_0\over N}[G({\bf n}),M_x;\lambda]=(1-M_x^{2})\Big[ D 
+ {1\over 2}\sum_{\bf n} \Delta({\bf n})
G({\bf n}) \Big] + \lambda \Big[ 1- {(1-M_x^2)}^{1\over2} n_{12} \Big]
\,.
\eqno(23)
$$
It turns out that the spin-exchange strength is dependent on $G$ and
$M_x$.  In the paramagnetic phase, the order parameter $M_x$ vanishes
identically and $E_0/N$ becomes a functional only of $G$. 

In mean-field theory, $G({\bf n}) \equiv 0$ and $n_{12}=1$, so
in this case $E_0$ becomes a function of $M_x$ only:
$$
{E_0\over N}(M_x)=(1-M_x^{2})D
+ \lambda \Big[ 1- {(1-M_x^2)}^{1\over2} \Big] \,. \eqno(24)
$$
This function is minimized by $M_x=\Big[ 1 -
(\lambda / 2D )^2 \big]^{1\over2}$, implying a critical point
at $\lambda_c = 2 D$, beyond which $M_x \equiv 0$.  The resultant
optimal energy is given by $E_0/N = D$ in the disordered (paramagnetic)
phase ($\lambda > 2D$) and by $E_0/N = \lambda (1- \lambda / 4D)$
in the ordered (ferromagnetic) phase ($0 \leq \lambda \leq 2D$).

Turning to the choice of variational wave function, much of the
physics of the transverse Ising model can be captured
by a correlated trial ground state of Hartree-Jastrow form:
$$
| \Psi_{\rm HJ} \rangle = \exp ( M_x U_{M_x}+U ) | 0 \rangle \,, \eqno(25)
$$
with
$$
U= {1\over 2}\sum_{i<j}^{N}u({\bf r}_{ij}) {\sigma}_{i}^{x}{\sigma}_{j}^{x}\,,
\eqno(26)
$$
$$
U_M=\sum_i^N u_1({\bf r}_i) \sigma_i^x + {1\over 4}\sum_{i<j}^{N}
u_M({\bf r}_{ij}) ({\sigma}_{i}^{x}+ {\sigma}_{j}^{x}) \,.
\eqno(27)
$$
The vacuum or reference state $| 0 \rangle $ is taken as a symmetric
product of $N$ single-spin states with spin components $\sigma_i^z = +1$,
i.e.
$$
| 0 \rangle = | \uparrow \uparrow \cdot \cdot \cdot \uparrow \rangle_N
\,.
\eqno(28)
$$
For a translationally invariant system, the single-spin function $u_1({\bf r}_i)$ 
is independent of the lattice site ${\bf r}_i$, while the two-body
pseudopotentials $u$ and $u_M$ depend only on the relative distance 
${\bf n}$, where ${\bf n} = {\bf r}_{ij} = {\bf r}_i - {\bf r}_j$.  
In mean-field approximation, $u({\bf n}) \equiv u_M({\bf n}) \equiv 0$.

In the disordered phase, the $U_M$ generator is not present in the
exponential form defining the correlated trial ground state,
since the order parameter $M_x$ vanishes identically.  However, this 
term makes a vital contribution in the ordered phase, where it 
is responsible for the symmetry breaking.  (Note that reflection 
in a mirror plane normal to the $x$-axis transforms $U_M$ to $-U_M$ 
and reveals a two-fold degeneracy of the ordered ground state
(characterized by the magnetizations $M_x$ and $-M_x$).  It should
also be mentioned that the pseudopotential $u_M({\bf n})$ is 
in fact a functional of the generator $u({\bf n})$.

Evaluation of the energy functional $E_0[G({\bf n},M_x;\lambda)]$
requires construction of the spatial distribution function $G({\bf n})$ and
the spin-exchange strength $n_{12}$ corresponding to the trial ground
state, as functionals of the pseudopotential $u({\bf n})$ that generates
the spatial correlations.  This is done efficiently by exploiting 
a $1-1$ mapping of the spin-lattice system onto a binary mixture 
of two boson species, made possible by the assumed form of the 
trial ground state.  The two boson species are characterized by eigenvalues 
$+1$ and $-1$ of the spin operator $\sigma^x$ and may be called 
{\sl particles} and {\sl holes}, respectively.  The partial densities 
$\rho_+$ and $\rho_-$ of particle and hole components are determined 
by the magnetization $M_x$ through $\rho_{\pm}={1\over 2}(1 \pm M_x)$, 
i.e., by the expectation values $ \rho_{\pm}= \langle P_{i}^{(\pm)} 
\rangle$ of the projectors $P_{i}^{(\pm)}={1\over 2}(1\pm \sigma_i^x)$.
The Hypernetted-Chain (HNC) analysis available for the Hartree-Jastrow 
 ground state of the binary boson mixture [31]
may then be applied to determine all the requisite quantities
for the corresponding variational description of the transverse Ising
system.

The trial ground state is optimized by deriving and solving suitable
Euler-Lagrange equations that determine the optimal distribution function 
$G({\bf n})$ the magnetization $M_x$.   For constant $\lambda$ and $M_x$, 
the optimal $G({\bf n})$ is determined, through its HNC connection
to the pseudopotential $u({\bf n})$, by means of the Euler-Lagrange
equation $\delta E_0 / \delta u({\bf n}) = 0$ for $u({\bf n})$,
which leads to a paired-magnon equation [24].  Similarly, variation
of the energy functional with respect to $M_x^2$ at constant
$\lambda$ and fixed $G({\bf n})$ produces an Euler-Lagrange 
$ \partial E_0 / \partial M^2=0 $ equation for the optimal order 
parameter in the ordered phase (moot in the disordered phase), which
leads to a renormalized Hartree equation [24].

This approach yields exact results in strong and weak-coupling limits, and
good results in between, but it cannot be expected to reproduce 
critical exponents without the inclusion of higher-spin correlations.
\vskip 28truept

\centerline{\bf 4.  ENTANGLEMENT IN HARTREE-JASTROW GROUND STATES}
\vskip 12 truept

Numerical data are available from published [23-25] and unpublished 
variational-CBF calculations in the transverse Ising model that 
suffice for meaningful evaluations of measures of corresponding 
bipartite entanglement properties for two-, three-, and 
four-dimensional ($D=2,3,4$) versions of the model.
\vskip 12truept

\noindent
{\sl Single-Site Entropy.}  Eqs.~(14) and (15) are used to quantify
the entanglement of a single site with the rest of the lattice (the
single-spin von Neumann entropy defined in Section II), using 
the variational-CBF inputs for $M_x$ and $M_z$ for many
choices of the coupling parameter $\lambda$.  The results, plotted 
in Fig.~1, indicate a sharp peaking of $S(\lambda)$ at the critical 
values of $\lambda$ for the order-disorder transition from ferromagnetism 
to paramagnetism given by the many-body calculation (respectively,
$\lambda_c = 3.14$, 5.12, and 7.1 for $D = 2$, 3, and 4).
It is tempting to interpret this peaking, with the entanglement 
measure reaching a maximum at the transition, in terms of 
a direct association of entanglement with quantum critical
phenomena.  The same behavior was observed by Osborne and
Nielsen for $D=1$, however with a distinctly higher maximum
value of $S$ (0.68 in comparison with the value 0.22 we
find at $D=2$).  In fact, the maxima are seen to decline 
systematically as $D$ increases.  This finding is in harmony
with the general understanding that classicality increases
with dimension.  This suggests that in practical implementations
of quantum information processing where entanglement is used
as a resource, it is advantageous to utilize chains of processing
units rather than arrays in higher dimensions.
\input psfig.sty
\topinsert
\vskip -5truecm
\psfig{figure=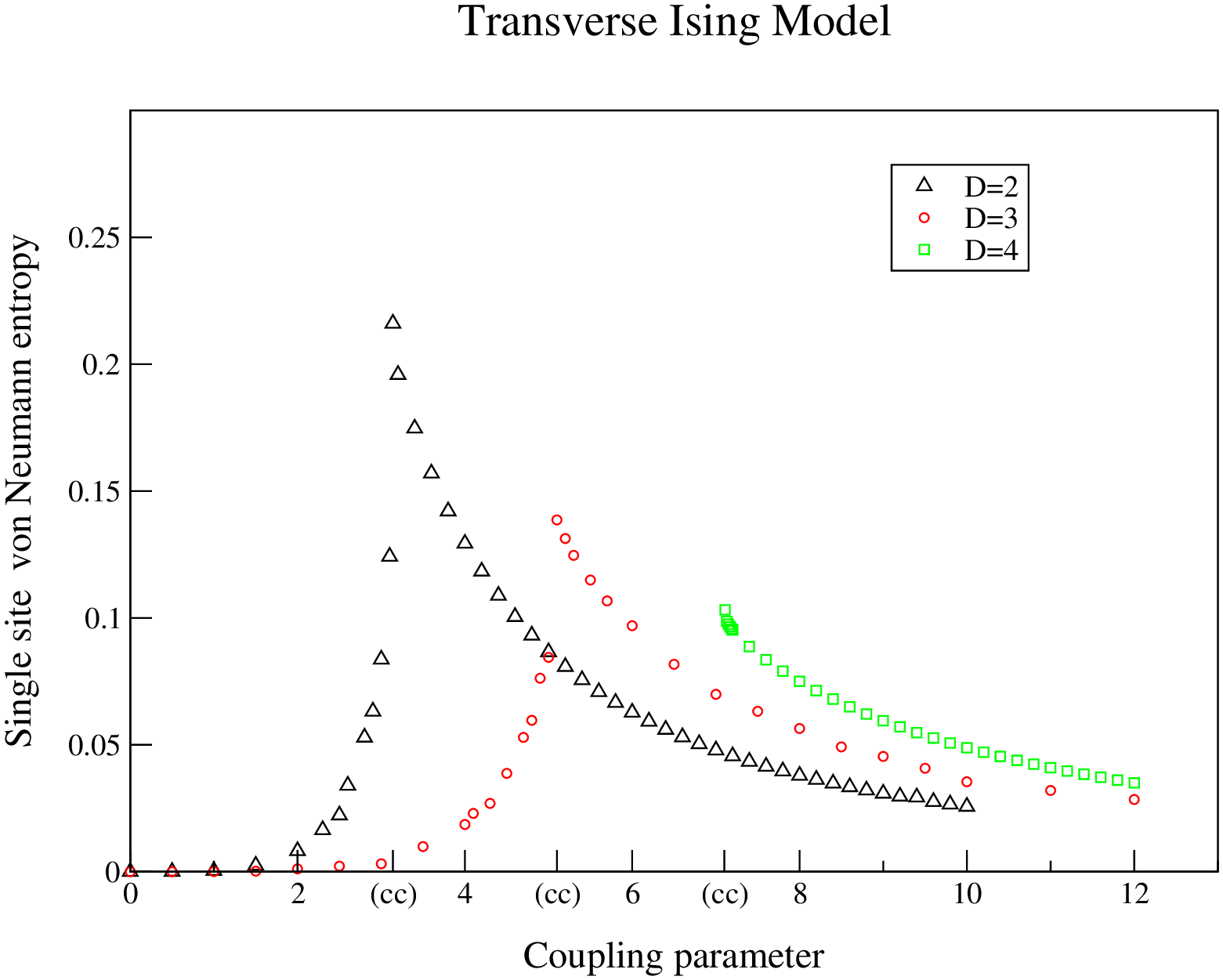,width=12truecm,angle=0}
\noindent
{\bf Figure 1.}
Von Neumann entropy $S$ between a single site and
the remaining sites in a square, cubic, or
hypercubic lattice (respectively for dimensions $D=2$, 3, or 4),
evaluated with input data from variational-CBF calculations
based on optimized Hartree-Jastrow ground-state trial functions.
The locations of the respective critical coupling parameters
$\lambda_c$ are labeled with (cc).
\vskip .4truecm
\endinsert

\noindent
{\sl Two-Site Entanglement.}  We next study the entanglement between
two spins residing in the lattice in terms of the concurrence defined
in Section II, deriving information on these measures from the data 
available from the CBF-variational studies based on the optimized 
Hartree-Jastrow trial function.  In applying Eq.~(4), We may use the 
following formulas for the eigenvalues $\mu_i$ of the matrix 
$R$ of Eq.~(5) in terms of spin-spin correlation functions, 
which are valid in both ordered and disordered phases 
$$
\mu_{1,2} = {1 \over 4} \left| 1 -  
\langle \sigma_i^z \sigma_j^z \rangle \pm                     
\left (\langle \sigma_i^x \sigma_j^x \rangle  +                  
\langle \sigma_i^y \sigma_j^y \rangle  \right)  \right| \,, 
$$
$$
\mu_{3,4} = {1 \over 4} \left| \left[ \left(1 + 
\langle \sigma_i^z \sigma_j^z \rangle\right)^2 -
\langle \sigma_i^z + \sigma_j^z  \rangle^2 \right]^{1/2}
\pm \left( \langle \sigma_i^x \sigma_j^x \rangle -
\langle \sigma_i^y \sigma_j^y \rangle  \right) 
\right| \,.  \eqno(29)
$$
\vskip .2truecm
\input psfig.sty
\topinsert
\vskip -3.5truecm
\psfig{figure=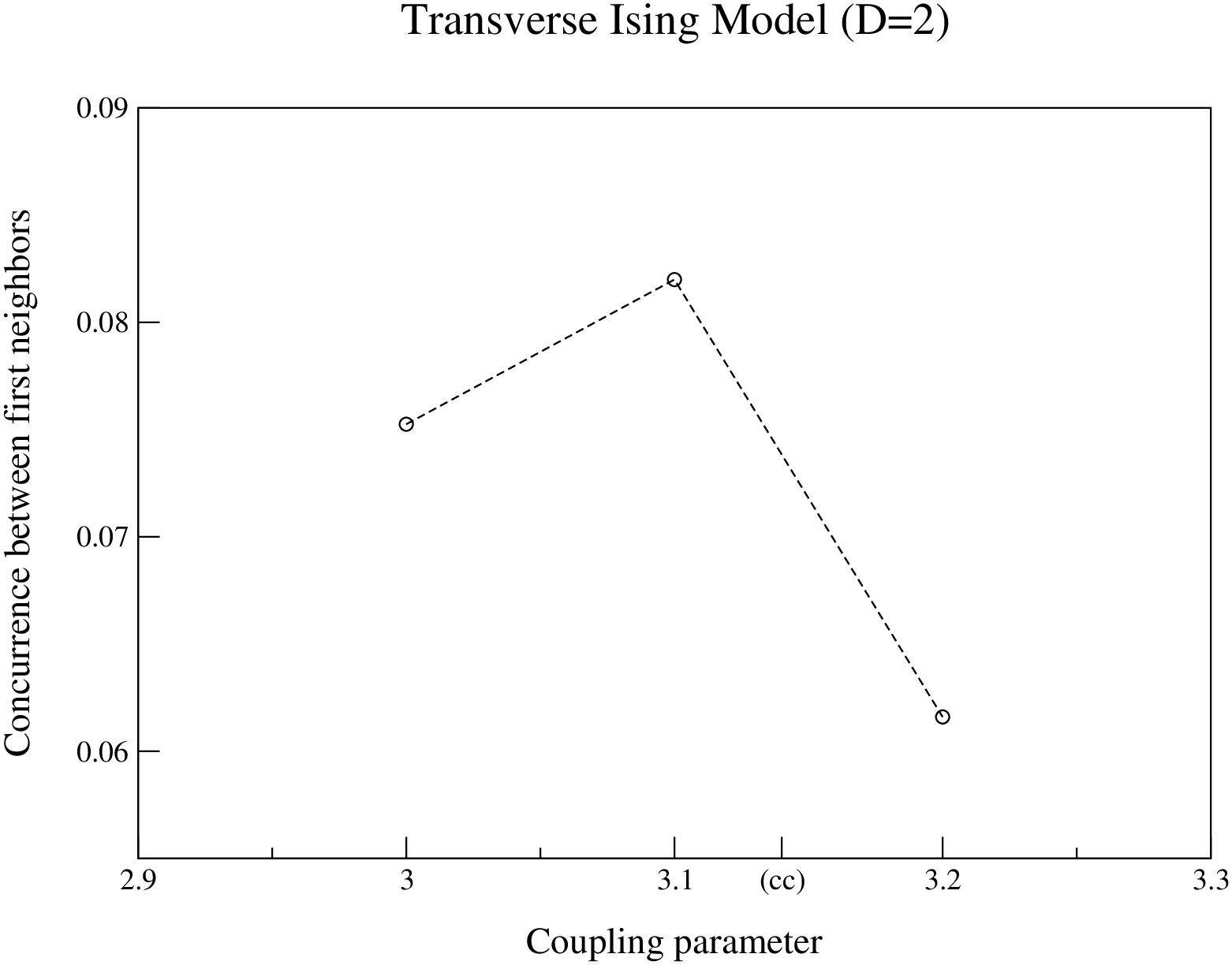,width=12truecm,angle=0}
\noindent
{\bf Figure 2.}
Concurrence $C$ between nearest neighbors in a square lattice,
evaluated with input data from variational-CBF calculations
based on an optimized Hartree-Jastrow ground-state trial function.
The location of the critical coupling parameter $\lambda_c$ is
labeled with (cc).  The dashed line serves merely to guide the eye.
\vskip .4truecm
\endinsert
\noindent
Concurrence results for nearest-neighbor spins in the two-dimensional
model are plotted in Fig.~2 at three values of the coupling parameter 
$\lambda$.  Since the existing data on the correlation functions 
is quite limited, this figure is not very informative.  Still, it is
of interest to point out that the values obtained are lower than
those obtained by Sylju{\aa}sen [28] based on Monte Carlo simulation
at finite temperatures, but the qualitative features are the same 
as those found in that work and by Osborne and Nielsen [27] and 
Osterloh et al.\ [26].  In our case, the peak may be closer to
the critical $\lambda$ than is the case for the chain [27].  As 
noted in Ref.~[28], one should expect the nearest-neighbor
concurrence to be smaller in higher dimension due to the monogamous 
character of entanglement -- the more neighbors, the smaller the 
share of bipartite entanglement allotted to each pair.  Also, we know 
quite explicitly from the results of ON and Osterloh et al.\ that 
although spin-spin correlations in the usual sense acquire a 
long-range character upon approach to the critical point, this 
is {\it not} the case for the concurrence.  In particular, these
authors find that $C$ vanishes for site separations on the chain beyond 
next-nearest neighbor.  In our case, since the scale of $C$ is
already smaller because of the higher dimensionalities considered, 
the concurrence is found to vanish for pair separations beyond 
nearest neighbor.
\vskip 12truept

\noindent
{\sl Localizable Entanglement.}  As indicated in Section II, 
localizable entanglement is another recently suggested measure of 
bipartite entanglement [22].  It has the virtue of providing for 
a more natural definition and incisive definition of entanglement 
length than the entanglement of formation, and allowing one
to observe its expected divergence at the quantum critical point.
For the case of the transverse Ising model, the $x$-connected
spin-spin correlation function $Q_{xx}({\bf n})$, furnishes a
tight lower bound on the localized entanglement, at least when
nearest-neighbor spins are considered [22].   

\input psfig.sty
\topinsert
\vskip -4truecm
\psfig{figure=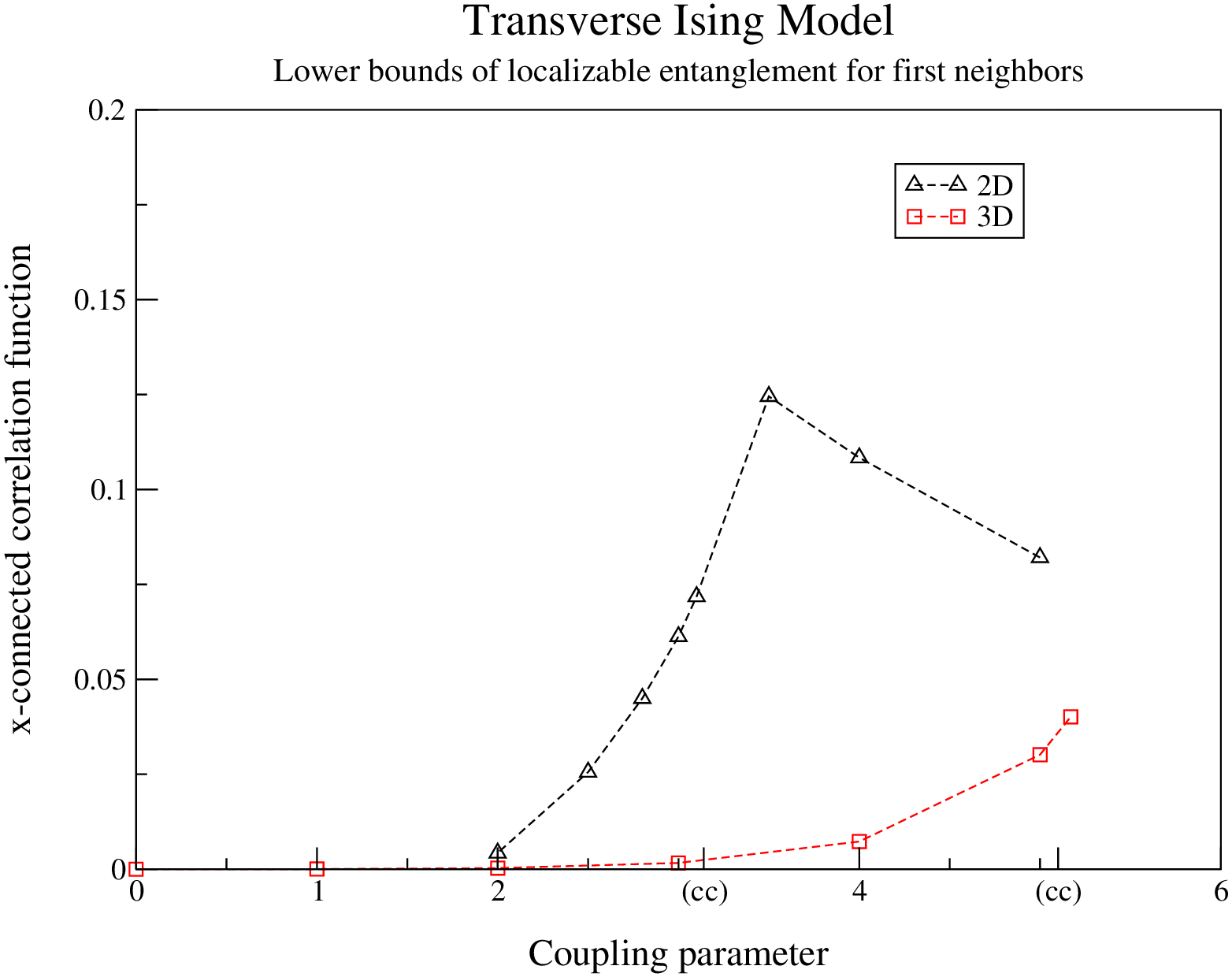,width=12truecm,angle=0}
\noindent
{\bf Figure 3.} 
Two-point connected correlation function $Q_{xx}(|{\bf n}|)$
for nearest neighbors on a square or cubic lattice (respectively 
for dimensions $D=2$ or 3), evaluated with input data from 
variational-CBF calculations based on Hartree-Jastrow 
ground-state trial functions.  The locations of the respective 
critical coupling parameters $\lambda_c$ are
labeled with (cc).  The dashed line serves merely to guide the eye.
\vskip .3truecm
\endinsert
Based on the theoretical analyses and numerical data of the
variational-CBF work reported in Refs.~[23-35], we have evaluated
$Q_{xx}(|{\bf n}|) = g(|{\bf n}|)-M_x^2$ for a number of 
$\lambda$ values, in the cases $D=2$ and 3 (see Fig.~3).  (Here,
$|{\bf n}|$ denotes the spin-spin distance in the lattice at
constant transverse magnetic field.)  It is important to add 
that the data used are not fully optimal.  If we compare with 
the corresponding results of Verstraete et al.\ [22],
it is seen that the variational results for $Q_{xx}$ behave in 
the same way as when, in their treatment, a small perturbing 
magnetic field is imposed in the $x$ direction to break the 
parity symmetry (i.e., under $x \to -x$) and achieve a more 
realistic description in the strong-coupling limit (small 
$\lambda$).  The treatment on which our input data is based 
already incorporates parity breaking in the ferromagnetic phase.
\vskip 12truept

\noindent
{\sl Tanglemeter}.
There have been some initial efforts toward analysis of
multipartite entanglement in quantum spin models, focusing on
spin chains [32-36].  In this vein, but not restricted to one 
dimension, we shall make some observations on the entanglement 
properties of Hartree-Jastrow ground states within the 
recently developed multipartite formalism based on nilpotent 
polynomials [9].

Given the ground state Eqs.~(25)-(27) determined by the 
pseudopotentials $U$ and $U_M$, one can easily
find the corresponding nilpotent polynomial $F$ defined
in Ref.~[9].  For the present
argument it is irrelevant whether or not $U$ and $U_M$ are
optimized.  It is very important that $U$ and $U_M$ are
respectively of two-body and one-body structure.  Thus, the
local contributions from $U_M$ can be ignored, since they do not
affect the entanglement properties of the state.  The required
construction leading to the so-called tanglemeter is further simplified
by the fact that the different terms and factors in $U$ commute
with one another.  The polynomial $F$ is then found to be
$$
F=1+{\sum}_{i<j} \alpha_{ij} \sigma_i^+\sigma_j^+ + {\sum}_{i<j<k<l} 
\alpha_{ijkl} \sigma_i^+\sigma_j^+\sigma_k^+\sigma_l^++ 
\ldots+\alpha_{12\ldots N}\sigma_1^+\sigma_2^+\ldots\sigma_N^+ \,,
$$
which does not contain terms linear or of order $N-1$ in the
$\sigma^+$ operators.  Therefore $F$ is already in its canonical
form $F_c$ under general local transformations, and the number
of the coefficients is less than the minimum number of parameters
sufficient to describe the entanglement properties of the state.
The last step is to determine the logarithm of the polynomial $F$,
the so-called nilpotential $f$.  The result for $f$ is of the same 
form as $F$ and so already in canonical and hence tanglemeter. 

From this simple analysis, we can infer that for any non-zero 
choice for the pseudopotential $U$ in the form (26), the ground
state (25) belongs to the general orbit of states
which contain $N$-partite entanglement.  However, this state
cannot be the GHZ state containing maximal $N$-partite entanglement.
Another implication is that there are no subclusters of spins that
are not entangled with the rest of the lattice.  All spins share
quantum correlations with all others.
\vskip 28truept

\centerline{\bf 5.  ENTANGLEMENT IN OTHER MANY-BODY WAVE FUNCTIONS}
\vskip 12 truept

Although much can still be learned from studies of entanglement
in lattice spin systems, the characterization of this fascinating
nonlocal quantal correlation in systems of particles having 
continuous {\it spatial} degrees of freedom as well as discrete 
spin/isospin observables presents the greatest opportunities for 
new insights relevant to strongly interacting quantum systems.

A typical CBF state has the form of a correlation factor applied to
a reference independent-particle model state.  The example 
of the transverse Ising model shows that a Jastrow factor 
introduces nontrivial entanglement properties.  It is instructive 
to compare the simple Jastrow ground state for a normal Fermi 
system, namely $\prod_{i<j}f(ij) \Phi$ where $\Phi$ is a Slater 
determinant, with the BCS ground state projected on the subspace 
with the same even number $N$ of particles, which has the 
form ${\cal A}\left(\chi(12) \chi(34)\cdot \cdot \cdot \chi(N-1, N) 
\right)$ (antisymmetrized by ${\cal A}$).  The BCS wave 
function is evidently separable, being a mean-field description 
of Cooper pairs all in the two-body state $\chi$.  Thus the useful 
entanglement is concentrated entirely in the individual Cooper pairs, 
which are not entangled one with the other.  As demonstrated
by Shi [19], the entanglement of each member of a Cooper pair 
with the other is given by a von Neumann entropy
$$
S = |v_k|^2 \log |v_k|^2 - (1-|v_k|^2) \log (1 - |v_k|^2) \,, \eqno(30)
$$
where $v_k$ is the amplitude appearing in the usual expression
$$
| {\rm BCS} \rangle
= \prod_k \left( u_k + v_k a_{k \uparrow}^\dagger
a_{-k \downarrow }^\dagger \right) | 0 \rangle \eqno(31)
$$
for the BCS state vector,
with $|v_k|^2 = (1/2)(1 - \epsilon_k/E_k )$ and
$E_k = (\epsilon_k + \Delta_k^2)^{1/2}$ in the usual
notation.  The pair-member entanglement is zero when the gap $\Delta_k$
vanishes (normal state), and the Cooper pairs are maximally
entangled when $E_k >> \epsilon_k$.

By contrast, within the Jastrow product of pair functions 
$f$, entanglement is spread among {\it all} $N(N-1)/2$ pairs $ij$, 
both directly and indirectly.  Inclusion of interparticle 
correlations of all orders (between pairs, triples, 
quartets, etc.) is accomplished with the Feenberg function 
$\exp\left[ \sum_{i<j} u_2(ij)/2 + \sum_{i<j<k} u_3(ijk)/2 + \cdots 
\right]$, which in itself can provide an exact representation of the
ground state of a system of indistinguishable bosons [37].  A study of the
multipartite entanglement properties of this function should prove 
very informative.  (Such an analysis may be naturally extended to
shadow wave functions [38], a generalization of the Jastrow-Feenberg
form used predominantly in the study of quantum fluids and solids.) 

For similar reasons, the exponential structure of the states of
coupled-cluster theory [39],  
$$
|\Psi_{\rm CC} \rangle = \exp \left[ S_1 + S_2 + \cdots S_N  \right]\,,
\eqno(32)
$$
wherein
$$
S_1 = \sum_{ij} c_{ij} a_i^\dagger a_j\,,\quad\quad
S_2 = \sum_{ijkl} c_{ijkl} a_i^\dagger a_j^\dagger a_k a_l\,,~ {\rm etc.}
\eqno(33)
$$
is a propitious format for the investigation of multipartite 
entanglement in many-body problems in nuclear physics and related 
areas, already begun by Emary [40].  Moreover, the CC structure 
suggests that a fruitful analysis can be performed in terms 
of the ``tanglemeter'' developed by Mandilara et al.\ [9] as 
an extensive characterization of entanglement based on nilpotent 
polynomials.  
\vskip 28truept

\centerline{\bf ACKNOWLEDGMENTS}
\vskip 12 truept
The research described herein was supported in part by the U.S.\ National 
Science Foundation under Grant No.~PHY-0140316. JWC is grateful to 
the US Army Research Office--Research Triangle Park for travel 
support through a grant to Southern Illinois University--Carbondale,
which permitted him to take part in CMT29 at Kizu.  He expresses
sincere appreciation to the Japanese hosts of the the Workshop for their 
splendid organization, generosity, and hospitality.  JWC and AM would
also like to acknowledge partial support from FCT POCTI, FEDER in 
Portugal and the hospitality of the Centro de Ci\^ncias Mathem\'aticas
of the University of Madeira in connection with Madeira Math 
Encounters XXIX, during which much of this work was done.
\vskip 28 truept

\centerline{\bf REFERENCES}
\vskip 12 truept

\item{[1]}
V. Vedral, M. B. Plenio, K. Jacobs, and P. L. Knight, {\it Phys.~Rev.~A}
{\bf 56}, 4452 (1997); V. Vedral, B. Plenio, M. A. Rippin, and
P. L. Knight, Phys. Rev. Lett. {\bf 78}, 2275 (1998).
\item{[2]}
N.~Linden, S.~Popescu, and A.~Sudbery, {\it Phys.~Rev.~Lett.} {\bf 83},
243 (1999); H. A. Carteret, N. Linden, S. Popescu, and A. Sudbery,
Found.~Phys. {\bf 29}, 527 (1999).
\item{[3]}
W. D\"ur, G. Vidal, and J. I. Cirac, {\it Phys.~Rev.~A} {\bf 62},
062314 (2000).
\item{[4]}
V. Vidral, {\it Rev. Mod. Phys.} {\bf 74}, 197 (2002).
\item{[5]}
J. Eisert, K. Audenaert, and M. B. Plenio, {\it J.~Phys.~A:~Math.~Gen.} 
{\bf 36}, 5605 (2003).
\item{[6]}
A. Miyake, {\it Phys. Rev. A} {\bf 67}, 012108 (2003).
\item{[7]}
F. Verstraete, J. Dehaene, and B. De Moor, {\it Phys. Rev. A}
{\bf 68}, 012103 (2003).
\item{[8]}
H. Barnum, E. Knill, G. Ortiz, R. Somma, and L. Viola,
{\it Phys. Rev. Lett.} {\bf 92}, 107902 (2004);
R. Somma, G. Ortiz, H. Barnum, E. Knill, and L. Viola,
{\it Phys. Rev. A} {\bf 70}, 042311 (2004).
\item{[9]}
A. Mandilara, V. M. Akulin, A. V. Smilga, and L. Viola, Description
of quantum entanglement with nilpotent polynomials, 
quant-ph/0508234 (2005), to be submitted to Phys.~Rev.~A. 
\item{[10]}
E.~Schr\"odinger, {\it Proceedings of the Cambridge Philosophical Society}
{\bf 31}, 555 (1935).
\item{[11]}
J. S. Bell, {\it Rev. Mod. Phys.} {\bf 38}, 447 (1966).
\item{[12]}
M. A. Nielsen and I. L. Chuang, {\it Quantum Computation and Quantum
Information} (Cambridge University Press, Cambridge, 2000).
\item{[13]}
J. Preskill, {\it Lecture Notes on Quantum Computation},
http://www.theory.caltech.\break edu/people/preskill/ph229/\#Lecture.
\item{[14]}
J. W. Clark and M. L. Ristig, {\it Theory of Spin Lattices and
Lattice Gauge Models}, Springer Lecture Notes in Physics, 
Vol.~494 (Springer, Berlin, 1997).
\item{[15]}
R.~F. Bishop, D.~J.~J.~Farnell, M.~L.~Ristig, 
{\it Int.~J.~Mod.~Phys.} {\bf 14}, 1517 (2000).  
\item{[16]} J. Schliemann, D. Loss, and A. H. McDonald, 
{\it Phys. Rev. B} {\bf 63}, 085311 (2001); 
J. Schliemann, J. I. Cirac, M. Ku\'s, M. Lewenstein, and 
D. Loss, {\it Phys. Rev. A} {\bf 64}, 022303 (2001);
K. Eckert, J. Schliemann, D. Bruss, and M. Lewenstein,
{\it Ann. Phys.} (N.Y.) {\bf 299}, 88 (2002).
\item{[17]}
R. Paskauskas and L. You, {\it Phys. Rev. B} {\bf 64}, 042310 (2001).
\item{[18]}
Y. S. Li, B. Zeng, X. S. Liu, and G. L. Long, {\it Phys. Rev. A}
{\bf 64}, 054302 (2001).
\item{[19]}
Y. Shi, {\it J. Phys. A: Math. Gen.} {\bf 37}, 6807 (2004).
\item{[20]}
C. H. Bennett, H. J. Bernstein, S. Popescu, and B. Schumacher,
{\it Phys. Rev. A} {\bf 53}, 2046 (1996).
\item{[21]}
W. K. Wootters, {\it Quantum Inf. Comput.} {\bf 1}, 27 (2001);
S. Hill and W. K. Wootters, {\it Phys. Rev. Lett.} {\bf 78}, 5022
(1997); W. K. Wootters, {\it Phys. Rev. Lett.} {\bf 80}, 2245 (1998).
\item{[22]}
F. Verstraete, M. Popp, and J. I. Cirac, {\it Phys. Rev. Lett.} {\bf 92},
027901 (2004); M. Popp, F. Verstraete, M. A. Mart\'in-Delagado, 
and J. I. Cirac, {\it Phys. Rev. A} {\bf 71}, 042306 (2005).
032110 (2002).
\item{[23]}
M. L. Ristig and J. W. Kim, {\it Phys. Rev. B} {\bf 53}, 6665 (1996). 
\item{[24]}
M. L. Ristig, J. W. Kim, and J. W. Clark, in {\it Theory of Spin
Lattices and Lattice Gauge Models} (Springer-Verlag, Berlin, 1997),
p. 62.
\item{[25]}
J. W. Kim, M. L. Ristig, and J. W. Clark, {\it Phys. Rev. B} {\bf 57}, 
56 (1998).
\item{[26]}
A. Osterloh, L. Amico, G. Falci, and R. Fazio, {\it Nature} {\bf 416}, 
608 (2002).
\item{[27]}
T. J. Osborne and M. A. Nielsen, {\it Phys. Rev. A} {\bf 66},
\item{[28]}
O.~F.~Sylju{\aa}sen, {\it Phys. Lett.} {\bf A322}, 25 (2004). 
\item{[29]}
J. W. Clark and E. Feenberg, {\it Phys.~Rev.}~{\bf 113}, 388 (1959);
J. W. Clark and P. Westhaus, {\it Phys.~Rev.}~{\bf 141}, 833 (1966);
E. Feenberg, {\it Theory of Quantum Fluids} (Academic Press, New York,
1969); C. E. Campbell and E. Feenberg, {\it Phys. Rev.} {\bf 188}, 
396 (1969); J. W. Clark, in {\it The Many-Body Problem, Jastrow 
Correlations versus Brueckner Theory}, Springer Lecture Notes in Physics, 
Vol.\ 138, R. Guardiola and J. Ros, eds.\ (Springer-Verlag, Berlin), p. 184;
C. E. Campbell, in {\it Recent Progress in Many-Body Theories}, Vol.~4
E. Schachinger, H. Mitter, and H. Sormann, eds.\ (Plenum, New York, 1995),
p. 29; E. Krotscheck, in {\it Lecture Notes in Physics}, Vol.\ 510,
J. Navarro and A. Polls, eds.\ (Springer, Heidelberg, 1998).
\item{[30]}
C. E. Campbell, K. E. K\"urten, M. L. Ristig, and G. Senger,
{\it Phys. Rev. B} {\bf 30}, 3728 (1984); G. Senger, M. L. Ristig,
K. E. K\"urten, and C. E. Campbell, {\it Phys. Rev. B} {\bf 33}, 7562
(1986); K. A. Gernoth, J. W. Clark, and M. L. Ristig, 
Z. Physik B {\bf 98}, 337 (1995).
\item{[31]}
M. L. Ristig, S. Fantoni, and K. E. K\"urten, {\it Z. Phys. B} {\bf 51}, 
1 (1983).
\item{[32]}
X. Wang, {\it Phys. Rev. A} {\bf 66}, 044305 (2002).
\item{[33]}
P. Stelmachovic and V. Buzek, {\it Phys.~Rev.~A} {\bf 70}, 032313 (2004).
\item{[34]}
A. Lakshminarayan and V. Subrahmanyam, quant-ph/0409039 (2005)
\item{[35]}
D. Bruss, N. Datta, A. Ekert, L. C. Kwek, and C. Macchiavello,
quant-ph/0411080 (2005).
\item{[36]}
O.~Guhne, G.~Toth, and H. J. Briegel, quantph/0502160 (2005).
\item{[37]}
J. W. Clark, {\it Nucl. Phys.} {\bf A328}, 587 (1979).
\item{[38]}
L. Reatto and G. L. Masserini, {\it Phys. Rev. B} {\bf 38}, 4516 (1988).
\item{[39]}
F. Coester, {\it Nucl.~Phys.} {\bf 7}, 421 (1958); F.~Coester and H.~K\"ummel, 
ibid.\ {\bf 17}, 477 (1960); J. Cizek, {\it J. Chem. Phys.} {\bf 45}, 4256 
(1966); {\it Adv.\ Chem.\ Phys.} {\bf 14}, 35 (1969); R.~F.~Bishop and 
K.~H.~L\"uhrmann, {\it Phys.\ Rev.} B {\bf 17}, 3757 (1978); H.~K\"ummel, 
K.~H.~L\"uhrmann, and J.~G.~Zabolitzky, {\it Phys.\ Rep.} {\bf 36C}, 1 (1978); 
J.~S.~Arponen, {\it Ann.~Phys.~(N.Y.)} {\bf 151}, 311 (1983); R.~F.~Bishop 
and H.~K\"ummel, {\it Phys.~Today} {\bf 40}(3), 52 (1987); J.~S.~Arponen, 
R.~F.~Bishop, and E.~Pajanne, {\it Phys.~Rev.~A} {\bf 36}, 2519 (1987); 
ibid.\ {\bf 36}, 2539 (1987); R.~J.~Bartlett, {\it J.~Phys.~Chem.} {\bf 93}, 
1697 (1989); R.~F.~Bishop, {\it Theor.\ Chim.\ Acta} {\bf 80}, 95 (1991). 
\item{[40]}
C. Emary, invited talk presented at the Conference on Microscopic
Approaches to Many-Body Theory, University of Manchester, Aug 31-Sept 3,
2005; and private communication.
\end
\noindent
{\bf Figure 1.}
Von Neumann entropy $S$ between a single site and
the remaining sites in a square, cubic, or
hypercubic lattice (respectively for dimensions $D=2$, 3, or 4),
evaluated with input data from variational-CBF calculations
based on optimized Hartree-Jastrow ground-state trial functions.
The locations of the respective critical coupling parameters 
$\lambda_c$ are labeled with (cc).  
\vskip 1truecm
{\bf Figure 2.}
Concurrence $C$ between nearest neighbors in a square lattice,
evaluated with input data from variational-CBF calculations
based on an optimized Hartree-Jastrow ground-state trial function.
The location of the critical coupling parameter $\lambda_c$ is
labeled with (cc).  The dashed line serves merely to guide the eye.
\vskip 1truecm
{\bf Figure 3.} 
Two-point connected correlation function $Q_{xx}(|{\bf n}|)$
for nearest neighbors on a square or cubic lattice (respectively 
for dimensions $D=2$ or 3), evaluated with input data from 
variational-CBF calculations based on Hartree-Jastrow 
ground-state trial functions.  The locations of the respective 
critical coupling parameters $\lambda_c$ are
labeled with (cc).  The dashed line serves merely to guide the eye.